\documentstyle[prb,aps]{revtex}
\newcommand{\doublespace}{
  \renewcommand{\baselinestretch}{1.75}
  \large\normalsize}
\begin{document}
\doublespace
\title{Effects of microscopic strain distribution on Ga$_{1-x}$In$_x$As
quantum wires grown by strain-induced lateral ordering}

\author{Liang-Xin Li, Jingze Sun and Yia-Chung Chang}
\address{Department of Physics and Materials Research Laboratory 
\\ University of Illinois at Urbana-Champaign, Urbana, Illinois 61801}

\date{\today}
\maketitle

\begin{abstract}

Band structures and optical matrix elements of
quantum wires (QWR's) made of short-period superlattices (SPS)
with strain-induced lateral ordering (SILO)
are investigated theoretically via an
effective bond-orbital model (EBOM) combined with a valence-force field (VFF)
model. Valence-band anisotropy, band mixing, and effects due to 
local strain distribution at the atomistic level are all taken into account.
In particular, Ga$_{1-x}$In$_x$As
QWR's grown by SILO process are considered.
A VFF model is used to find the equilibrium atomic
positions in the SILO QWR structure by minimizing the lattice energy.
The strain tensor at each atomic (In or Ga) site is then obtained
and included in the calculation of electronic states and optical properties. 
It is found that different local arrangement of atoms leads to very different
strain distribution, which in term alters the optical properties.
In particular, we found
that the optical anisotropy can be reversed due to the change in shear
strain caused by the inter-change of atomic positions. Good agreement
with the existing experimental data on band gap and optical anisotropy
can be obtained when a 2D alloy structure with lateral composition modulation 
in the InAs/GaAs interface planes of the SPS is used. Our studies revealed
the possibility of "shear-strain engineering" in SILO QWR light-emitting
devices to achieve desired optical anisotropy.

\end{abstract}

\mbox{}

\mbox{}

\newpage

\section{Introduction}

Optical properties of III-V semiconductor nanostructures have attracted 
a great deal of interest, as they
are important for applications in optical communications that 
involve switching, amplification, and signal processings.
Ga$_x$In$_{1-x}$As and Ga$_x$In$_{1-x}$P 
are among   the most important ternary III-V 
compound semiconductors. Ga$_x$In$_{1-x}$As in particular has band gaps 
covering both the 1.3 and 1.55 $\mu m$
range, which are the preferred wavelengths in long distance fiber 
communications[1-3]. However, long-wavelength photonic devices based on 
lattice-matched Ga$_{0.47}$In$_{0,53}$As/InP heterostructures suffer 
from strong Auger recombination and intervalence band absorption 
processes[4-7].
Recently, to improve the performance of long-wavelength semiconductor 
lasers, long wavelength($\sim 1.55 {\mu}$m) Ga$_x$In$_{1-x}$As quantum-wire 
(QWR) lasers have been grown by a single step molecular beam 
epitaxy technique[1].

An important optical property is the change in the emission of light 
(in energy, polarization, 
and intensity) that results from phase-space filling of carriers in 
one- and two-dimensionally confined 
systems, i.e., As the dimensionality of the quantum confinement increases 
from one dimension (1D) to 2D, 
the narrowing of the density of states will exhibit a lower excitation 
threshold for phase-space filling, 
thereby, yielding potentially enhanced optical effects.
It is found that the QWR laser structures are a
promising candidate for optical communication device because of 
the many predicted benefits, such as higher gain, 
reduced temperature sensitivity, higher modulation bandwidths, and narrower 
spectral linewidths[2]. 

The fabrication of quantum wires via the strain-induced lateral-layer ordering
(SILO) process starts with the growth of
short-period superlattices (SPS) [e.g. (GaAs)$_2$/(InAs)$_{2.2}$]
along the [001] direction. The excess fractional InAs layer leads to
stripe-like islands during the MBE growth. The continued growth of
(GaAs)$_2$/(InAs)$_{2}$ SPS on top of this surface will lead to
spatial separation of Ga-rich stripes and In-rich stripes due to the
effect of strain potential.[1,2] Thus, a strain-induced lateral ordering occurs
and a quantum wire heterostructure can then be
created by simply utilizing this SPS structure with lateral
composition modulation as the quantum well region in a 
conventional quantum heterostructure. 
In the experiment performed by	Chou et al[1] and Tang et al[2],
the QWR active region is created 
$\it{in~ situ~}$ by the SILO process within the (GaAs)$_2$/(InAs)$_2$  
SPS regions. The SILO process generates a strong Ga/In lateral 
composition modulation and creates Ga-rich $Ga_xIn_{1-x}As$ lateral QWs 
in the [110] direction. By sandwiching the composition modulated layer
between Al$_{0.24}$Ga$_{0.24}$In$_{0.52}$As barrier layers,
a strained QWR heterostructure is formed[1,2]. 
Besides the self-assembled lateral
ordering, it is believed that the strain also plays a key role[2] 
in the temperature stability and optical anisotropy for the QWR laser structure.
Recent studies indicate that the use 
of the strained-layer quantum wire heterostructure has advantages of 
high-quality interfaces and band-gap tuning which are 
independent of the lattice constant of the 
constituent materials[3,4]. Furthermore, much work has been undertaken which 
predicts that by using strained-layer superlattice to form the active 
region of a 
quantum-wire laser, the threshold current can be decreased by one order of 
magnitude, and the optical loss due to intervalence-band absorption and  Auger 
recombination will also be greatly reduced[1,2,5-8]. Also, the temperature 
sensitivity 
is reduced by an order of magnitude compared with strain-free structures[2].
A typical 
temperature sensitivity of the lasing wavelength is($\sim 5 \AA/{}^0C$) for 
the usual 
GaInAs/InP lasers. By using a distributed-feedback structure, the temperature 
dependence of the lasing wavelength is reduced to $1\AA/{}^0C$[2]. 
With this strained 
GaInAs QWR, the dependence is smaller than $0.1\AA/{}^0C$[2]. It should be an 
important improvement on the current technology in fabricating the 
long wavelength lasers for fiber communication. 
In this paper we study the effects of multi-axial strain on
the electronic and optical properties of the QWR structures grown via 
the SILO process. Through this study, we can gain a better 
understanding of the strain engineering of  
QWR structures suitable for the application of fiber-optical communication.

	In a previous paper[9], we performed theoretical calculations of
the optical properties of SILO QWR's within the virtual crystal approximation
in which the SPS region is modeled by a Ga$_x$In$_{1-x}$As alloy with a lateral
modulation of the composition $x$. Although our calculation can explain
the QWR band gap and optical anisotropy approximately, it does not take into account the
detailed SPS structure and the microscopic strain distribution.
The understanding of these effects is important if one wish to have a full 
design capability of the SILO QWR optoelectronic devices. 

In the present paper, we consider two possible local atomic arrangements
with SPS structure in the SILO quantum wires and examine the effects of
the microscopic strain distribution on the electronic and optical properties. 
The QWR model structures considered in the present paper are depicted
in Fig. 1. Both model structures start with a quantum well nanostructure
in which the supercell consists
of 8 stacks of (001) (GaAs)$_2$/(InAs)$_2$ SPS
with a total thickness of $\approx 100 \AA$ (quantum well region)
followed by a Al$_{0.24}$Ga$_{0.24}$In$_{0.52}$As layer ( barrier region)
with thickness $\approx 60 \AA$ (20 diatomic layers).
In the first QWR model structure (a), the strain-induced lateral ordering is
modeled by interchanging a segment of 29 Ga atoms in
the right half of the first row in the SPS (each row in
the SPS consists of 72 atoms along the [110] direction such that the
period is around 300 \AA)
with a segment of 29 In atoms in the left half of the third row. [see Fig. 1(a)]
This structure is an idealized model of a SILO QWR which starts with
a growth of  (GaAs)$_2$/(InAs)$_{2.2}$ SPS. It is assumed that during growth
an extra 0.4 monolayer of In atoms on the surface form stripe-like islands 
aligned along the [$1\bar 1 0$] direction [corresponding to the segment of 29 
In atoms in the first row of Fig. 1(a)], which serves as a seed structure 
for the formation of the SILO QWR. The deposition of two more GaAs monolayers
leaves extra 0.4 monolayer of Ga atoms on the surface, which again
form stripe-like islands [corresponding to the segment of 29 Ga atoms
in the third row of Fig. 1(a)]. Due to the strain potential, the Ga stripes
are spatially separated from the In stripes buried two monolayers
beneath.  Thus, the continued growth of
(GaAs)$_2$/(InAs)$_2$ SPS on top of the seed structure will lead to
a SILO QWR structure shown in Fig. 1(a).

In the second QWR model structure (b), we assume that the extra 0.4 monolayer
of In atoms on the surface of the seed structure mix randomly with
the newly deposited Ga atoms during growth so as to form a composition
modulated 2D alloy structure on the surface. The In composition $x$ is 
assumed to
vary in the [110] direction (or $y'$ direction) according to the relation
[see Fig. 1(c)]
\begin{equation}
x=  \left\{ \begin{array}{ll}	0  & \mbox{ for } b < y' < L_1-b \\
 0.4 + 0.4 \sin [\pi (y'-L_1)/2b] &  \mbox{ for } L_1-b < y' < L_1+b \\
 0.8   &	  \mbox{ for } L_1+b < y' < L-b \\
 0.4 - 0.4 \sin [\pi (y'-L)/2b] &  \mbox{ for } y'<b \mbox{ or }
L-b < y', \end{array} \right.
\end{equation}
where $2b$ denotes the width of composition grading, $L_1$ is the
length of Ga-rich region in the QWR, and $L$ is the period of the
lateral modulation in the [110] direction.
Here we use $b=6 a_{[110]} \approx 20 \AA$(??).

In both model structures, we can divide the SILO QWR into
two regions with the left half being Ga rich and the right half In rich.
The average Ga (In) mole fraction in the Ga (In)-rich region is 0.7 (0.3).
Alternatively, we can divide the first structure into three regions
which include (InAs)$_1$(GaAs)$_3$, (InAs)$_2$(GaAs)$_2$,
and (InAs)$_3$(GaAs)$_1$ SPS's with the In mole fraction changes abruptly
between boundaries of the three regions. In the second structure, the
In mole fraction for the whole QWR varies continuously between 0.3
and 0.7 for $y'$ changing from the Ga-rich region to the In-rich region.

A valence force field (VFF) model[10-12] is used to find the equilibrium 
atomic positions in the SILO QWR structure by minimizing the lattice energy.
The strain tensor at each atomic (In or Ga) site is then obtained by
calculating the local distortion of chemical bonds.
We found that different local arrangement of atoms can lead to very different
strain distribution. In particular, the shear strain can change
substantially from structure 1 (with abrupt change in In composition)
to structure 2 (with gradual change in In composition).
Furthermore, we found
that the optical anisotropy can be reversed due to the change in the
strength of the shear strain caused by
the intermixing of In and Ga atoms.
We found that both structures give band gaps close to experimental value.
However, only the second structure gives good agreement
with the published experimental data[1,2] on both band gap and optical 
anisotropy, indicating that random mixing of In and Ga atoms on the 
growth surface when the island structure is present best matches
the experimental situation.

\section{Theoretical Approach}
The method used in this paper for calculating the strained QWR band 
structure is based 
on the effective bond-orbital model (EBOM). A detailed description of this 
method has been published elsewhere[13,14].
EBOM is a tight binding-like model with minimum set of localized basis
(the bonding or anti-bonding orbitals).
The interaction and optical parameters are obtained by a correspondence
with the ${\bf k\cdot p}$ theory, which can be cast into analytic forms. 
Thus, the model can be viewed as a spatially
discretized version of the ${\bf k\cdot p}$ method, while retaining the 
virtues of LCAO (linear combination of atomic orbitals) method.
The ${\bf k\cdot p}$ model is the most popular one for treating electronic
structures of semiconductor quantum wells or superlattices.
However, when applied to complex structures such as SILO quantum wires[1-4]
or self-assembled quantum dots[10,15], the method becomes very cumbersome
if one wish to implement the correct boundary conditions that
take into account the differences in
${\bf k\cdot p}$ band parameters for different materials involved.
EBOM is free of this problem, since different material parameters are
used at different atomic sites in a natural way. For simple structures, when
both EBOM and ${\bf k\cdot p}$ model are equally applicable,
the results obtained are essentially identical.[13]

The optical matrix elements for the QWR states are computed
in terms of elementary optical matrix elements between the valence-band
bond orbitals and the conduction-band orbitals. 
The present calculation includes the coupling of the four spin-3/2 
valence bands and the two spin-1/2 conduction bands closest to the band edges.
Thus, it is equivalent to a 6-band ${\bf k\cdot p}$ model.
For our systems studied here, the band-edge properties are relatively 
unaffected by the split-off band due to the large spin-orbit splitting
as discussed in our previous paper[9]. Hence, the split-off bands are 
ignored here. The bond-orbitals for the GaAS and InAs	needed in
the expansion of the superlattice states contain the following: 
four valence-band bond orbitals per bulk unit cell, which are p-like 
orbitals coupled with the spin to form orbitals with total 
angular momentum J=3/2 plus two conduction-band bond orbitals with J=1/2. 
They are written as
\begin{equation}
|{\bf R},u_{JM}>=\sum_{\alpha,\sigma}C(\alpha,\sigma,J,M)|\vec{R},\alpha>
\psi_{\sigma},
\end{equation}
where $J=1/2$ and 3/2 for the conduction and valence bands, respectively, 
and $M=-J,\cdots J$, $\psi_{\sigma}$, designates the electron 
spinor($\sigma$=1/2,-1/2), and $|{\bf R},\alpha>$ denote an 
$\alpha$-like($\alpha=s,x,y,z$) bond orbital, located at unit cell 
${\bf R}$. $C(\alpha,\sigma,J,M)$ are the coupling coefficients 
obtainable by group theory. All these bond orbitals are assumed 
to be sufficiently localized so that the interaction between orbitals 
separated farther than the nearest-neighbor distance can be ignored. 

    The effect of strain is included by adding a strain Hamiltonian 
$H^{st}$ to the EBOM Hamiltonian[14].  The matrix elements 
of $H^{st}$ in the bond-orbital basis can be 
obtained by the deformation-potential theory of Bir and Pikus[16]. 
We use the valence-force field (VFF) model of Keating and Martin[11,12]
to calculate the microscopic strain distribution. This model has been shown 
to be successful in fitting and predicting the elastic constants of 
elastic continuum theory, calculating strain distribution in a quantum well, 
and determining the atomic structure of III-V alloys. 
It was also used in the calculation of the strain distribution in 
self-assembled quantum dots.[10] The VFF model is a microscopic theory 
which includes bond stretching and bond bending, and avoids the potential 
failure of elastic continuum theory in the atomically thin limit. 
The total energy of the lattice is taken as 

\begin{equation}
V =\frac{1}{4}\sum_{ij}\frac{3}{4}\alpha_{ij}(d^2_{ij}-d^2_{0.ij})^2/d^2_{0,ij}
+\frac{1}{2}\sum_i\sum_{j\neq k}\frac{3}{4}\beta_{ijk}(\vec{d_{ij}}.\vec{d_{ik}}
+d_{0,ij}d_{0,ik}/3)^2/d_{0,ij} d_{0,ik}
\end{equation}

\noindent where i runs over all the atomic sites, j, k run over the 
nearest-neighbor sites of i, $\vec{d_{ij}}$ is the vector joining the 
sites i and j, $d_{ij}$ is the length of the bond, $d_{0,ij}$ is the 
corresponding equilibrium length in the 
binary constituents, and $\alpha_{ij}$ and $\beta_{ijk}$ are the 
bond stretching and bond bending constants, respectively. 
$\alpha$ and $\beta$ are from Martin's calculations.[11] 
For the bond-bending
parameter $\beta$ of In-As-Ga, we take $\beta_{ijk}=\sqrt{\beta[ij]\beta[ik]}
$ following Ref. [17]. 
     
  To find the strain tensor in the $InAs/GaAs$ SILO QWR, we start from  
ideal atomic positions and minimize the system energy 
using the 
Hamiltonian given above. Minimization of the total energy requires one to 
solve a set of coupled equations with 3N variables, where N is the total 
number of atoms. Direct solution of these equations is impractical 
in our case, since the system contains more than 6,000 atoms. 
We use an approach taken by several authors which has been shown to be 
quite efficient. In the beginning of the simulation all the atoms are 
placed on the InP lattice, we allow atoms to deviate from this starting 
positions and use periodic boundary conditions in the plane
perpendicular to the growth direction, while keeping atoms in the
planes outside the SPS region at their ideal atomic positions
for a InP lattice (since the SILO QWR is grown epitaxially on the InP 
substrate). In each iteration, only one atomic 
position is displaced and other atom positions are held fixed. 
The direction of the displacement of atom $i$ is determined according to
the force  $f_i=\partial V/\partial x_i$ acting on it.
All atoms are displaced in sequence. the whole sequence is
repeated until the forces acting on all atoms become zero
at which point the system energy is a minimum.
Once the positions of all the atoms are known, the strain distribution 
is obtained through the strain tensor calculated according to the method
described in Ref. [18].

Let $R^0$ be the position matrix without strain:
\[
R^0  = \left ( \begin{array}{ccc}
       R^0_{12x} & R^0_{23x} & R^0_{34x} \\
       R^0_{12y} & R^0_{23y} & R^0_{34y} \\
       R^0_{12z} & R^0_{23z} & R^0_{34z}  
      \end{array} \right ), \]
where ${\bf R}^0_{ij} = {\bf R}^0_j - {\bf R}^0_i; i,j=1,4$ and ${\bf R}^0_i$
$(i=1,4)$ denote positions of four As atoms surrounding a Ga or In atom.
Here we choose ${\bf R}^0_{12} = (1, -1, 0)a/2$, 
${\bf R}^0_{23} = (-1, 0, -1)a/2$, and ${\bf R}^0_{34} = (1, 1, 0)a/2$. 
$a$ is the lattice constant of GaAs or InAs, depending on the site.
Let $R$ be the corresponding position matrix with strain. It was shown
in Ref. [18] that the strain tensor 
is given by

\begin{equation}
\epsilon=R*R_0^{-1} - 1.
\end{equation}

After getting the strain tensor, the strain Hamiltonian is given by Bir and
Pikus[16]

\begin{equation}
H^{st} = \left( \begin{array}{ccc} 
 -\Delta V_H + D_1 &\sqrt{3}d e_{xy} &\sqrt{3}d e_{xz} \\
  \sqrt{3}d e_{xy} & -\Delta V_H + D_2     &\sqrt{3}d e_{yz} \\ 
       \sqrt{3}d e_{xz} &\sqrt{3}d e_{yz} & -\Delta V_H + D_3
               \end{array} \right),
\end{equation}
where $e_{ij}=(\epsilon_{ij}+\epsilon_{ji})/2$, and
$$
\Delta V_H = (a_1+a_2)(\epsilon_{xx}+\epsilon_{yy}+\epsilon_{zz}),~~\\
 D_1 = b(2\epsilon_{xx}-\epsilon_{yy}-\epsilon_{zz}),~~\\
 D_2 = b(2\epsilon_{yy}-\epsilon_{xx}-\epsilon_{zz}),~~\\
 D_3 = b(2\epsilon_{zz}-\epsilon_{yy}-\epsilon_{xx}),\\
$$
The strain potential on the s states is given by
$$\Delta V_c = c_1(\epsilon_{xx}+\epsilon_{yy}+\epsilon_{zz}),$$
The strain Hamiltonian in the bond-orbital basis $|JM>$ can be easily 
found by using the coupling constants[13], i.e, 
\begin{equation}
<JM|H^{st}|J'M'>=\sum_{\alpha, \alpha',\sigma}C(\alpha, \sigma;J,M)^*
C(\alpha', \sigma;J',M')H^{st}_{\alpha\alpha'}
\end{equation}
The elastic constants $C_{12}$ and $C_{11}$ for GaAs, InAs and AlAs can be found
in Ref. [18,19]. The deformation potentials $a_1,~a_2,~b,~c_1, ~d$ can be
found in Ref.[20,21]. The linear interpolation
and virtual crystal approximations used to obtain the corresponding
parameters for the barrier material (Al$_{0.24}$Ga$_{0.24}$In$_{0.52}$As).

The above strain Hamiltonian is derived locally for the each cation atom in 
the SILO QWR considered. To calculate the electronic states of the SILO QWR,
we first construct a zeroth-order
Hamiltonian for a superlattice structure which contains in each period
8 stacks of (001) (GaAs)$_{2}$/(InAs)$_{2}$ SPS layers
(with a total thickness around 100 \AA) and 20 diatomic layers of 
Al$_{0.24}$Ga$_{0.24}$In$_{0.52}$As (with thickness around 60 \AA).
So, the superlattice unit cell for the zeroth-order model contains
52 diatomic layers.  The appropriate strain Hamiltonian for the
the(GaAs)$_{2}$/(InAs)$_{2}$ SPS  on InP is also included.

The eigen-states for the zero-th order Hamiltonian for different
values of $k_2$ (separated by the SL reciprocal lattice vectors
in the [110] direction) are then used as the basis
for calculating the SILO QWR electronic states.
The difference in the Hamiltonian (including strain effects)
caused by the intermixing of Ga and In
atoms at the interfaces is then added to the zeroth-order Hamiltonian,
and the electronic states of the  full Hamiltonian is solved by 
diagonalizing the Hamiltonian matrix defined within a truncated set of
eigen-states of the zeroth-order Hamiltonian. A total of $\sim 300$ eigenstates
of the zero-th order Hamiltonian (with 21 different $k_{110}$ points)
were used in the expansion. The subbands
closest to the band edge are converged to within 0.1 meV.

\section{Results and discussions}

{\bf A. Strain distributions}

In this section we discuss strain distributions in two QWR model structures
as described in section I. In both structures composition modulation is along 
the $y'$ ([110]) direction. In structure 1 atomic species are uchanged along 
$x'$ ([1,$\bar 1$,0]), while in structure 2 we have an
2D In$_x$Ga$_{1-x}$As alloy with the composition $x$ varying
as a function of $y'$. 
The diagonal and shear strains of structure 1 are shown in Figs. 2 and 3,
respectively. For best illustration, we show diagonal strains  
in a rotated frame, in which $x'$ is [1,$\bar 1$,0], $y'$ is [110], 
and  $z'$ is [001].  Shear strains are shown in
the original Cartesian coordinates.

	Since the atomic species does not change along $x'$, the $x'$ coordinate
of all atoms are very close to the original values pinned by the substrate
(InP), thus $\epsilon_{x'x'}$ (solid lines) are essentially constants 
along $y'$ in regions where atomic species are same.
For example, Layer 2 contains all In atoms, and $\epsilon_{x'x'}$
is constant with a value equal to  $(a_{InAs}-a_{InP})/a_{InAs}$.
In layer 1, Ga atoms occupy the sites labeled y=40-68, and the rest are
In atoms. Thus, there is a discrete jump of $\epsilon_{x'x'}$
from $a_{InP}-a_{InAs}/a_{InAs}$ to $a_{InP}-a_{GaAs}/a_{GaAs}$.
In the ideal situation (without lateral relaxation), $\epsilon_{y'y'}$
should be the same as $\epsilon_{x'x'}$ due to symmetry. However, with
the lateral modulation as considered here, all Ga(In) atoms tend to shift
in a direction so as to reduce the strain in the Ga(In)-rich region, while
they shift in the opposite direction in the In(Ga)-rich region. 
Thus, the magnitude of $\epsilon_{y'y'}$ on Ga(In) sites
(dashed lines in Fig. 2) is lower than $\epsilon_{x'x'}$ in
Ga(In)-rich region and vice versa in In(Ga)-rich region.
On the other hand, the $z$ component strain (dash-dotted lines) tend to 
compensate the other two components such that the volume of each bulk unit cell
is close to that for the unstrained bulk.
Thus, we see that $\epsilon_{z'z'}$  has an opposite sign
compared to $\epsilon_{x'x'}$ or $\epsilon_{y'y'}$ at all atomic sites. 

	The difference in $\epsilon_{x'x'}$ and $\epsilon_{y'y'}$ 
leads to a nonzero shear strain $e_{xy}=(\epsilon_{y'y'}-\epsilon_{x'x'})/2$
in the original coordinates. The shear strain is particularly strong at 
the interface where the atomic species changes abruptly. (see solid
lines in Fig. 3). The other shear strains $\epsilon_{xz}$ and
$\epsilon_{yz}$ (dash-dotted lines)
are also quite significant (around 1-2 \%) in structure 1.
The structure has a mirror plane normal to the $x'$ axis; thus 
we have $\epsilon_{xz}=\epsilon_{yz}$ here.

To model the alloy structure with composition modulation (structure 2),
we use a super-cell
which contains 72 atoms in the [110] ($y'$) direction, 36 atoms in the
[$1\bar 1 0$] ($x'$) direction and 18 atomic planes along the [001] ($z$)
direction [2 stacks of (2/2) SPS plus one monolayer of GaAs latticed
matched to InP].
In the atomic planes which consist of alloy structure, we first determine
the In composition at a given $y'$ according to Eq. (1) and then use a
random number generator to determine the atomic species along the
$x'$ direction. The calculated strain distributions are then averaged over
the $x'$ coordinate. The middle four layers are
used as the supercell for the SILO QWR structure as shown in Fig. 1(b).
Their strain distributions are shown below.

    Fig. 4 shows the diagonal strain distributions of the second
structures in the rotated coordinates ($x',y',z'$).
Comparing with those in structure 1, strain distributions are quite different 
in structure 2. Due to the random
distribution of In/Ga atoms along $x'$ in the alloy layers, 
the difference of $\epsilon_{x'x'}$ and $\epsilon_{y'y'}$ is
significantly reduced. 
Since In composition $x$ 
changes sinusoidally between Ga-rich and In-rich regions in layer 1 (3) 
$\epsilon_{x'x'}$ and $\epsilon_{y'y'}$  also
change gradually from their GaAs (InAs) bulk values to the alloy bulk values. 
The average values of $\epsilon_{x'x'}$ in layers 
1 and 3 are essentially the softened profiles of those shown in Fig. 2 for 
structure 1, while they remain approximately constant in 
layers 2 and 4. 

    Fig. 5 shows the off-diagonal strain distributions of the second
structures in the original coordinates. Due to the gradual change of In 
composition from Ga-rich
to In-rich regions, the shear strain is on average five times
smaller than its counterpart in structure 1. The largest $xy$-component shear
strain (solid line) is around 0.4\%, which occurs near the boundary
between Ga-rich and In-rich regions. 
We show below that the significant difference
in shear strains between the two QWR model structures will lead to dramatically
different optical properties. 

{\bf B. Electronic structures}

    In order to understand the aspect of lateral quantum confinement
due to lateral composition modulation and strain, we first examine
the band-edge energies of a strained quantum well structure whose
well material is the same as the SPS structure appeared in the SILO
QWR with a fixed value of $y'$ coordinate.
For structure 1 depicted in
Fig. 1(a), the well material consists of 8 stacks
(GaAs)$_{2-n}$/(InAs)$_{2+n}$ SPS with $n=-1, 0, 1$ in different
regions of the SILO QWR. In structure 2, the
(GaAs)$_{2-n}$/(InAs)$_{2+n}$ SPS is replaced by a
GaAs$_2$/Ga$_{1-x}$In$_{x}$As/InAs (GaAs/Ga$_{1-x}$In$_{x}$As/InAs$_2$)
SPS in the Ga-rich (In-rich) region of the SILO QWR with $x$ varying as
a function of $y'$ according to Eq. (1).

The results for the conduction
band minimum and valence band maximum of the above quantum well
as functions of $y'$
are shown in Fig. 6 (dashed lines). For comparison, the corresponding
band edges for a SPS structure without the
Al$_{0.24}$Ga$_{0.24}$In$_{0.52}$As confining barrier are also
shown with (dash-dotted) and without (solid) the effect of strain.
The strain Hamiltonian used here is the same as the one used in
the SILO QWR at the corresponding $y'$.
All material parameters are chosen
the same as in Ref. [9] for temperature at 77K, except that the deformation
potentials used here are slightly different. We use $C_1=-9.3,-4.1$, $a_1+a_2=
-2.7,-2.5 eV$ for GaAs and InAs, respectively. 
These values are within the uncertainties of 
experimentally determined values[20,21] and they give a better agreement 
for the band gaps of (GaAs)$_2$(InAs)$_2$ SPS  and SILO QWRs between
the calculated and experimental values. The photoluminecence (PL) measurements
indicates that the (GaAs)$_2$(InAs)$_2$ SPS grown on InP
substrate has a gap around 0.76 eV.[22] Her we obtain a band gap of
0.76 eV for the (GaAs)$_2$(InAs)$_2$ SPS (at $y'$=36) and
0.78 eV for 8 stacks (or 100 \AA) of (GaAs)$_2$(InAs)$_2$ SPS 
sandwiched between Al$_{0.24}$Ga$_{0.24}$In$_{0.52}$As confining barriers.
This is also consistent with the PL measurements on the
(GaAs)$_2$(InAs)$_2$/InP multiple quantum wells.[22] 
As seen in this figure, the strain effect causes a shift of
conduction band edge by $\sim 0.5$ eV  ( 0.4 eV)
and valence-band edge by  $\sim 0.4$ eV ( 0.3 eV)  in structure 1 (structure 2).
The band-edge profiles shown in this figure suggest that both electrons
and holes are confined in the Ga-rich region with a band offset
around 0.15 eV for the electron and 0.05 eV for the hole.
Both offsets are large enough
to give rise to strong lateral confinement for electrons and holes in
the Ga-rich region.

Fig. 7 shows the near zone-center valence subband structures of
the QWR model structure 1 as depicted in Fig. 1(a).
All subbands are two-fold degenerate at the zone center
due to the Kramer's degeneracy and
they split at finite wave vectors as a result of lack of
inversion symmetry in the system.
Comparing the band structures in both $k_1$ ($[1\bar 10]$) and $k_2$
($[110]$) 
directions, we noticed an apparent anisotropy in the energy dispersion.
The dispersions in the $k_2$ direction for the (confined) valence
bands are rather small, indicating strong lateral confinement.
The first three pairs of subbands are labeled V1, V2, and V3.
They have unusually large energy separations compared with other
valence subbands. This is because the first three pairs of subbands represent
QWR confined states, whereas the other valence subbands (with energies below
-110 meV) are unconfined by the lateral composition modulation.
To examine the effects caused by the shear strain, we have also calculated
the band structures with the shear strain set to zero. We found that
without the shear strain the first pairs of valence subbands (V1) have 
much larger dispersion along the $k_1$ direction with an effective masses
along $[1\bar 10]$ about a factor five smaller than the effective mass
for the V1 subbands with the shear strain. This indicates that the
V1 subband states for the case without shear strain are derived
mainly from bond orbitals with $x'$-character, which leads to stronger
overlap between two bond orbitals along the $x'$ direction, hence 
larger dispersion along $k_1$. When the shear strain is present, the character
of bond orbitals in the V1 subband states change from mainly $x'$-like 
to predominantly $y'$-like; thus, the dispersion along $k_1$ becomes
much weaker. This explains the very flat V1 subbands as shown in Fig. 7.
Physically, the switching of orbital character in the V1 subbands can be
understood as follows. When the  shear strain is absent, we have 
$\epsilon_{x'x'}=\epsilon_{y'y'}$, and the confinement effect in the $y'$
direction pushes down the subband with $y'$ character (which has smaller
effective mass in the $y'$ direction), thus leaving the top valence 
subband (V1) to have [predominantly $x'$ character. On the other hand,
with the presence of shear strain as shown in stricture 1, 
we have $\epsilon_{y'y'} > \epsilon_{x'x'}$ in Ga-rich region, 
and the strain potential forces the $y'$-like subband to move above
the $x'$-like subband, overcoming the confinement effect. As a result,
the V1 subbands become $y'$-like.
The conduction subbands are approximately
parabolic as usual with a zone-center subband minimum equal to
716 meV 
This gives an energy gap of 791 meV  for QWR model structure 1.

Fig. 8 shows the optical matrix elements squared ($P^2$) for transitions 
from the highest three pairs of valence subbands to the first pair of
conduction subbands. Here $P^2$ is defined
as $\frac 2 {m_0} 
\sum_{s,s'} |<\psi_{v}|\hat \epsilon \cdot {\bf p}|\psi_{c}>|^2$,
where $\psi_{v}$ ($\psi_{c}$) denotes a valance (conduction) suband state.
The symbol $\sum_{s,s'}$ means a sum over two near degenerate pair of
subbands in the initial and final states. $\hat \epsilon$ denotes
the polarization of light. Here we consider only the $x'$ (along the
QWR axis) and $y'$ (perpendicular to the QWR axis) components.
As seen in this figure, the C1-V1 transition for $y'$-polarization (dashed
line) is about two times that for the $x'$-polarization (solid line),
indicating an optical anisotropy
($(P^2_{[1\bar 1  0]}-P^2_{[110]})/(P^2_{[1\bar 1 0]}+P^2_{[110]})$) 
around 0.4.
This is consistent with the discussion for valence subband structures shown
in Fig. 7, where we concluded that the bond orbitals involved in the
V1 subbands are predominantly $y'$-like.
The optical anisotropy for QWR model structure 1 is reversed compared 
with the optical anisotropy observed in the
photoluminescence measurements for most InGaAs SILO QWRs.
As pointed out in our previous paper, the lateral confinement in a typical QWR
(without shear strain) leads to a positive optical anisotropy[9].
Here we found the opposite result. This apparent "anomaly"
is caused by the presence of the strong shear strain (especially the
$xy$ component) in the QWR model structure. If we
artificially turn off the shear strain, then
the normal (positive) optical anisotropy  can be resumed.
We have studied several other model structures.
We found that as long as we maintain an ordered structure along the $x'$
(with the same periodicity as bulk),
the lateral modulation in the $y'$ direction is necessarily a sharp
transition in In composition, and the structure always has a large
shear strain near the boundary where the In composition changes.
The fact that most SILO QWRs  display a positive optical
anisotropy indicating that these structures can not be modeled
by a ordered SPS structure as depicted in Fig. 1 (a). On the other hand,
there do exist a few SILO QWR samples that display a slightly
negative optical anisotropy at 77K[23]. Perhaps, these SILO QWR structures have
more abrupt change in In composition along the [110] direction.

    Knowing that a positive optical anisotropy can occur in a QWR structure
with weak shear strain, we now turn to our model structure 2, in which a
gradual lateral modulation in In composition substantially reduces the
shear strain (see Fig. 5). The calculated valence subband structures for
this QWR model structure are shown in Fig. 9. Comparing this figure with
Fig. 7, we find striking differences in band structures for the
two QWR model structures. The first pairs of subbands (labeled V1) in Fig. 9
has  much larger dispersion than their counterparts in Fig. 7, and there exists
an anti-cross behavior between the first pair of subbands (labeled V1) and
the lower subbands (similar to the LH1-HH2 anti-crossing pattern
typically appeared in III-V quantum wells[24,25]) near $k_1=0.02 (2\pi/a)$
in Fig. 9. This anti-crossing pattern is missing in Fig. 7 for
QWR model structure 1. Again, the presence of the strong shear strain
in the QWR structure is responsible for this difference. By turning off
the shear strain artificially in QWR model structure 1, we found that the
dispersion of the V1 subbands become similar to that shown in Fig. 9
with an anti-crossing pattern.
The conduction band minimum (not shown) is 676 meV for QWR model structure 2.
This gives rise to a band gap of 767 meV, which is very good agreement
with the experimental results for InGaAs SILO QWR with similar
specification.

Fig. 10 shows the squared optical matrix elements ($P^2$) versus wave vectors
of  QWR model structure 2 for transitions involving the
first three pairs of valence subbands (V1, V2, V3) and the first pair
of conduction subbands (C1). The solid and dashed lines are for
the polarization vector along the $k_1$ ($[1\bar 1 0]$) (parallel to
the wire) and $k_2$ ([110]) (perpendicular to the wire) directions.  
We found for the parallel polarization (solid lines),
the squared optical matrix element for the V1-C1 transition has a maximum at
the zone center with a value near 20 eV
and remains close to this value for all finite $k_2$ and for
$k_1< 0.01 (2\pi/a)$.
For $k_1$ from	$0.01$ to  $0.02 (2\pi/a)$, it drops very quickly
as a result of band mixing with lower valence subbands.	For the
perpendicular polarization, the optical strength for the
V1-C1 transition is nearly zero for
finite $k_2$ and $k_1=0$. It gradually increases as $k_1$ increases and
become appreciable for $k_1 > 0.02 (2\pi/a)$.
This means that the bond orbitals involved in the V1 subband states
are mostly $x'$-like. This is consistent with the consequence
of the lateral confinement on a HH1-like quantum well state (see discussions
in Ref. 9).  Comparing Fig. 10 with Fig. 8, we see that for the V1-C1
transition not only the optical anisotropy is reversed, but their $k_1$
dependence is also qualitatively different.  In Fig. 8, the optical
strength for the C1-V1 transition decays very slowly as $k_1$ increases,
showing no sign of band-mixing effect, consistent with the absence
of anti-crossing pattern in Fig. 7.

To calculate the optical anisotropy in the photoluminescence spectra, 
we integrate the squared optical matrix
element over the range of $k_1$ corresponding to the spread of exciton 
envelope function in the $k_1$ space. The exciton envelope function
is obtained by solving the 1D Schr\"{o}ding equation for the exciton in
the effective-mass approximation similar to what we did in Ref. 9.
The exciton binding energy obtained is 23 meV.
We found that the ratio of the averaged  
optical strength for the V1-C1 transition for the parallel to perpendicular
component of the polarization vector is 0.5 and 3.82  for the
model structures 1 and 2 considered here.
Our results for QWR model structure 2 are qualitatively similar to
that reported in Ref. 9 for a QWR structure made of 3D alloy structures
with composition modulation.
Here we have a 2D alloy structure with composition modulation and a SPS
structure in the third direction. Furthermore, we have included the
effects of microscopic strain distribution at the atomistic level.
The agreement with the
experimental observation for the band gap (735 meV) and ratio of parallel
to perpendicular optical strength (around 2 - 4 for a number of QWR
structures)[1,2] is also much better than that obtained in the 3D
alloy structure model.
Thus, we conclude that our model structure
2 has captured all the essential features of the
realistic experimental SILO QWR structure.

\section{Summary}

We have calculated the band structures and optical matrix elements 
for the strained GaInAs  QWR grown by the SILO method. The actual SPS
structure and the microscopic strain distribution has been
taken into account. The effects of microscopic strain  distribution
on the valence subband structures and optical matrix elements are discussed
for two model SILO QWR structures, one with ideal periodicity along the
wire axis and abrupt changes of In composition in the lateral direction,
while the other with a 2D alloy mixing and gradual lateral modulation of
the In composition in the interface layers of the SPS structure.
The valence force field (VFF) model is used to calculate the equilibrium
atomic positions in the model QWR structures. This allows the calculation
of the strain distribution at the atomistic level.

We found that in model structure 1 or any similar structures with
ideal periodicity along the wire axis, the strain distribution always
has large off-diagonal (shear strain) components, which can
alter the valence subband structures substantially and give rise
to a reversed optical anisotropy compared with QWR structures with
negligible shear strain (such as the model structure 2).
This points to a possibility of "shear strain-engineering" to obtain
QWR laser structures of desired optical anisotropy.

    The band gap and optical anisotropy obtained for model structure 2 are
in very good agreement with experimental observations on most InGaAs SILO
QWRs with similar specifications. This indicates that our model structure
2 is fairly close to the realistic structure at least in the essential physical
aspects. The optical anisotropy obtained in our model structure 1 is
revered compared with that in model structure 2 and in the photoluminescence
measurements for most InAs/GaAs SILO QWRs. However, there exist a few
InAs/GaAs SILO QWR samples which display the reversed optical anisotropy
at 77 K.[23] Furthermore, strained InAs/GaAs QWRs with ideal periodicity
along the [$1\bar 10$] axis may be grown via
e-beam lithography followed by {\em in situ} regrowth. 
It will be interesting to see
if these ideal QWR structures grown in the future will display the reversed
optical anisotropy as we predicted here.

\vspace{1cm}
\leftline{\bf ACKNOWLEDGEMENTS}
\vspace{1ex}
  This work was supported in part by the National Science Foundation (NSF) 
under Grant No. NSF-ECS96-17153. J. S. was also supported by
a subcontract from the University of Southern California under the
MURI program, AFOSR, Contract  No. F49620-98-1-0474.
 We would like to thank K. Y. Cheng
and D. E. Wohlert for fruitful discussions
and for providing us with the detailed experimental data
of the QWR structures considered here.

\newpage
\vspace{2cm}
Figure Captions

\vspace{1ex}

\noindent Fig. 1.  Schematic sketch of the basic (2,2) SPS segment in
the SILO quantum wire for two model structures considered in the present paper.
Each segment consists of four diatomic layers with 72 bulk unit cells in
each layers. In structure 1, 29 Ga atoms in layer 1 are inter-changed with
29 In atoms in layer 3 so that the supercell is divided into two regions with
the left being Ga-rich (region A) and the right being In-rich (region B).
In structure 2, the first layer and third layer consist of 2D alloy structure
with lateral modulation of the In composition as shown in the profile below.
The 2D alloy structure models the mixing of Ga and In atoms during crystal
growth when Ga or In islands of 0.4 monolayer are present.

\noindent Fig. 2. Diagonal strain distribution of the first SILO QWR structure
[corresponding to Fig. 1(a)] in rotated frame.
Dotted: $\epsilon_{x'x'}$. Solid: $\epsilon_{y'y'}$.
Dashed: $\epsilon_{zz}$.

\noindent Fig. 3. Off-diagonal strain distribution of the first SILO
QWR structure [corresponding to Fig. 1(a)] in original frame.
Solid: $e_{xy}$. Dashed: $e_{xz}$ Dotted: $e_{yz}$.

\noindent Fig. 4. Diagonal strain distribution of the second SILO QWR structure
[corresponding to Fig. 1(b)] in rotated frame.
Dotted: $\epsilon_{x'x'}$. Solid: $\epsilon_{y'y'}$.
Dashed: $\epsilon_{z'z'}$.

\noindent Fig. 5. Off-diagonal strain distribution of the second SILO
QWR structure [corresponding to Fig. 1(b)] in original frame.
Solid: $e_{xy}$. Dashed: $e_{xz}$ Dotted: $e_{yz}$.

\noindent Fig. 6. Conduction and valence band edges
of the constituent materials for two SILO QWR structures
with and without the effect of strain as functions of the
[110] coordinate $y'$. Solid lines are for unstrained SPS structure,
dashed lines are for the SPS structure under the same
$y'$-dependent strain distribution as in
the corresponding SILO QWR structure,
and dash-dotted lines are for 8 stacks of SPS structures
sandwiched between 60\AA Al$_{0.24}$Ga$_{0.24}$In$_{0.52}$As barriers
including the QWR strain distribution.

\noindent Fig. 7. Valance subband structures for the
SILO QWR structures depicted in Fig. 1(a).

\noindent Fig. 8. Squared optical matrix elements for transitions from
the top three pairs of valence subbands to the first pair of
conduction subband for light polarized parallel (solid) and
perpendicular (dashed) to the QWR axis for the SILO QWR
structure depicted in Fig. 1(a).

\noindent Fig. 9. Valance subband structures for the
SILO QWR structure depicted in Fig. 1(b).

\noindent Fig. 10. Squared optical matrix elements for transitions from
the top three pairs of valence subbands to the first pair of
conduction subband for light polarized parallel (solid) and
perpendicular (dashed) to the QWR axis for the SILO QWR
structure depicted in Fig. 1(b).
                  

\newpage

\begin{thebibliography}{wide label}

\bibitem{chou}S.T. Chou, K. Y. Cheng, L. J. Chou, and K. C. Hsieh, Appl. 
Phys. Lett. {\bf 17}, 2220 (1995); J. Appl. Phys. {\bf 78} 6270, (1995); 
J. Vac. Sci. Tech. B {\bf 13}, 650 (1995);
K. Y. Cheng, K. C. Hsien, and J. N. Baillargeon, Appl. Phys. Lett. {\bf 60}, 
2892 (1992).
\bibitem{wohlert}D. E. Wohlert, S. T. Chou, A. C Chen, K. Y. Cheng, 
and K. C. Hsieh, Appl. Phys. Lett. {\bf 17}, 2386 (1996).
\bibitem{tang}Y. Tang, H. T. Lin, D. H. Rich, P. Colter and S. M. Vernon, Phys. Rev. {\bf 53B} R10501, (1996). 
\bibitem{matthews} J. W. Matthews and A. E. Blakeslee, J. Cryst. Growth 
{\bf 27}, 18 (1974).
\bibitem{gourley}P. L. Gourley, J. P. Hohimer, and R. M. Biefeld, Appl. Phys. 
Lett. {\bf 47}, 552 (1985).
\bibitem{adams}A. R. Adams, Electron. Lett. {\bf 22}, 249 (1986).
\bibitem{yablonocitch}E. Yablonovitch and E. O. Kane. IEEE J. Lightwave 
Technol. LT-4, 504 (1986).
\bibitem{agrawa}G. P. Agrawa and N. K. Dutta, Long Wavelength Semiconductor
 Lasers, 2nd ed. (Van. Nostrand Reinhold, New York, 1993) Chap.7.
\bibitem{li}L. X. Li and Y. C. Chang, J. Appl. Phys. {\bf 84} 6162, 1998. 
\bibitem{jiang}H. Jiang and J. Singh, Phys. Rev. B{\bf 56}, 4696(1997).

\bibitem{martin}R. M. Martin, Phys. Rev. B{\bf 1}, 4005(1969).
\bibitem{keating}P. N. Keating, Phys. Rev. {\bf 145}, 637(1966).
\bibitem{chang}Y. C. Chang, Phys. Rev. B{\bf 37}, 8215 (1988).
\bibitem{houng}Mau-Phon Houng and Y.C Chang, J. Appl. Phys. {\bf 65}, 3096
(1989).
\bibitem{cusack}M.A.Cusack, P.R. Briddon, and M. Jaros, Phys. Rev. B{\bf 56}, 4047(1997).
\bibitem{bir}G. L. Bir and G. E. Pikus, Symmetry and Strain Induced Effects 
in Semiconductors (Halsted, United Kingdom, 1974); 
L.D. Landau and E. L. Lifshitz, Theory of Elasticity 
(Addison-Wesley Publishing Company Inc, Reading, Massachusetts, USA, 1970).
\bibitem{podqorny}M. Podgorny,M.T. Czyzyk, A. Balzarotti, P. Letordi, N. Motta, A. Kisiel, and M. Zimmal-Starnawska, Solid State Comm.{\bf 55}, 413(1985).
\bibitem{pryor}C. Pryor, J. Kim, L.W. Wang, A.J. Williamson and A. Zunger, Phys. Rev. B{\bf 183}, 2548(1998).
\bibitem{adachi}S. Adachi, J. Appl. Phys. {\bf 53}, 8775 (1982).
\bibitem{mathieu}H. Mathieu, P. Meroe, E. L. Amerziane, B. Archilla, J. 
Camassel, and G. Poiblaud, Phys. Rev. B{\bf 19}, 2209 (1979); 
S. Adachi and C. Hamaguchi, Phys. Rev. B{\bf 19}, 938 (1979). 
\bibitem{hinckley}J. M. Hinckley and J. Singh, Phys. Rev. B{\bf 42},
3546 (1990).
\bibitem{land}O. Madelung and M. Schulz, Landolt-Borstein (1982).
\bibitem{Eaz} M. Razeghi, Ph. Maurel, F. Omnes, and J. Nagle, Appl. Phys.
Lett. {\bf 51}, 2216 (1987).
\bibitem{wohlert1} D.E. Wohlert, and K. Y. Cheng, private communications. 
\bibitem{Sanders} G. D. Sanders and Y. C. Chang, Phys. Rev. B {\bf 31}, 6892
(1985).
\bibitem{Sanders1}G. D. Sanders and Y. C. Chang, Phys. Rev. B{\bf 45}, 
9202 (1992).






\end{thebibliography}
\end{document}